\documentclass[10pt,conference]{IEEEtran}

\IEEEoverridecommandlockouts
\usepackage{cite}
\usepackage{amsmath,amssymb,amsfonts}
\usepackage{algorithmic}
\usepackage{graphicx}
\usepackage{textcomp}
\usepackage{xcolor}
\usepackage{enumitem}
\usepackage{makecell}
\usepackage{threeparttable}
\usepackage{multirow}
\usepackage{fontawesome}
\usepackage{subfigure}
\usepackage{verbatim}
\usepackage{balance}
\usepackage{pifont}
\usepackage{adjustbox}
\usepackage{rotating}
\usepackage[hidelinks]{hyperref}
\usepackage{booktabs}
\usepackage{inconsolata}
\usepackage{tcolorbox}
\usepackage[lined,boxed,linesnumbered,commentsnumbered,ruled]{algorithm2e}
\usepackage{color}

\newcommand{\tech}{CodeDenoise}

\newcommand{\Comment}[1]{}

\def\BibTeX{{\rm B\kern-.05em{\sc i\kern-.025em b}\kern-.08em
    T\kern-.1667em\lower.7ex\hbox{E}\kern-.125emX}}

\begin{document}

\title{On-the-fly Improving Performance of Deep Code Models via Input Denoising}

\author{\IEEEauthorblockN{Zhao Tian}
\IEEEauthorblockA{\textit{College of Intelligence and} \\
\textit{Computing, Tianjin University}\\
China \\
tianzhao@tju.edu.cn}
\and
\IEEEauthorblockN{Junjie Chen$^\dag$}
\IEEEauthorblockA{\textit{College of Intelligence and} \\
\textit{Computing, Tianjin University}\\
China \\
junjiechen@tju.edu.cn}
\and
\IEEEauthorblockN{Xiangyu Zhang}
\IEEEauthorblockA{\textit{Department of Computer Science,} \\
\textit{Purdue University}\\
USA \\
xyzhang@cs.purdue.edu}
\thanks{$^\dag$Junjie Chen is the corresponding author.}
}

\maketitle


\begin{abstract}
Deep learning has been widely adopted to tackle various code-based tasks by building deep code models based on a large amount of code snippets.
While these deep code models have achieved great success, even state-of-the-art models suffer from noise present in inputs leading to erroneous predictions. 
While it is possible to enhance models through retraining/fine-tuning, this is not a once-and-for-all approach and incurs significant overhead. 
In particular, these techniques cannot on-the-fly improve performance of (deployed) models.
There are currently some techniques for input denoising in other domains (such as image processing), but since code input is discrete and must strictly abide by complex syntactic and semantic constraints, input denoising techniques in other fields are almost not applicable.
In this work, we propose the first input denoising technique (i.e., \tech{}) for deep code models. 
Its key idea is to localize noisy identifiers in (likely) mispredicted inputs, and denoise such inputs by cleansing the located identifiers.
It does not need to retrain or reconstruct the model, but only needs to cleanse inputs on-the-fly to improve performance. 
Our experiments on 18 deep code models (i.e., three pre-trained models with six code-based datasets) demonstrate the effectiveness and efficiency of \tech{}.
For example, on average, \tech{} successfully denoises 21.91\% of mispredicted inputs and improves the original models by 2.04\% in terms of the model accuracy across all the subjects in an average of 0.48 second spent on each input, substantially outperforming the widely-used fine-tuning strategy.

\end{abstract}


\section{Introduction}
\label{sec:introduction}
Over the years, deep learning (DL) has been extensively applied to solve tasks of software engineering, such as code clone detection~\cite{white2016deep} and functionality classification~\cite{zhang2019novel}.
Various deep learning models are built based on a large amount of training code snippets.
Following the existing work~\cite{le2020deep,tian2023adversarial}, we call them \textit{deep code models}.
These models have received extensive attention from both academia and industry due to their excellent performance on various tasks.
For example, advanced deep code models have been integrated into industrial solutions, such as GitHub Copilot~\cite{copilot2023}, to enhance software development productivity.

Despite the popularity, deep code models also suffer from the performance issue (i.e., mispredictions) like DL models in other domains (e.g., image processing~\cite{xie2019feature} and speech recognition~\cite{germain2018speech}).
These mispredictions can negatively affect the practical use of code models, even slow down the software development process and damage software quality.
To improve model performance, designing more advanced neural networks for retraining models or incorporating more data for fine-tuning models are the most widely-used strategies~\cite{xie2019feature,yefet2020adversarial,henke2022semantic,zhang2022toward}.
However, they cannot improve model performance on-the-fly after the models have been deployed.
Moreover, the retraining or fine-tuning process can be time-consuming caused by manual labeling and heavy computations, and also largely inexplicable for improving model performance.


As demonstrated by the existing work~\cite{lecun1989backpropagation,krizhevsky2017imagenet}, many mispredictions are actually caused by noise in inputs.
For deep code models, noise refers to 
perturbations in input code, which do not incur grammar errors and damage original code semantics~\cite{yefet2020adversarial}.
Intuitively, identifying and removing noise from mispredicted inputs can improve model performance on-the-fly.
This method is more \textit{efficient} and \textit{explicable} than model retraining/fine-tuning, and \textit{applicable to deployed models}.
Although some denoising methods have been proposed for images and speech via feature denoising neural networks, such as LRCnet~\cite{ren2022robust} and CleanUNet~\cite{kong2022speech}, they cannot be applied to deep code models.
This is because they denoise inputs in continuous space but the inputs (i.e., source code) for deep code models are discrete.
Moreover, source code has to strictly stick to complex syntactic and semantic constraints (i.e., the inputs after denoising should have no grammar errors and preserve the original semantics), which further aggravates the difficulty of denoising inputs.

In this work, we propose \tech{}, the first input denoising technique for deep code models, designed to on-the-fly improve model performance (especially for deployed models).
To design an effective input denoising technique, we require to address two challenges:
(1) how to \textit{identify} noise resulting in misprediction from a given code snippet;
(2) how to \textit{cleanse} noise to make the code snippet be predicted correctly.
To overcome the first challenge, \tech{} analyzes the contribution of each element in the code to the misprediction by utilizing the built-in attention mechanism~\cite{bahdanau2015neural} in advanced deep code models.
The code elements with larger contributions are more likely to be the noise resulting in the misprediction.
Here, \tech{} considers identifier-level noise since many existing studies have demonstrated that identifiers have a significant impact on deep code model performance~\cite{zhang2020generating,yang2022natural,zhang2022towards}.
To overcome the second challenge, \tech{} builds a {\em masked clean identifier prediction} (MCIP) model based on the typical masked identifier prediction task~\cite{feng2020codebert,wang2021codet5}, which aims to correct misprediction by replacing the identified noisy identifiers with the predicted clean ones according to the context. 
In this way, the code after denoising still has the same semantics as the original code. 

When deploying \tech{} in practice, we cannot know whether an incoming code snippet is predicted correctly or mispredicted by the code model, and thus cannot determine whether \tech{} should be triggered to denoise the code snippet.
To make \tech{} self-contained during the practical use, we design a strategy to estimate whether a code snippet is potentially mispredicted in \tech{}.
That is, if a code snippet is identified as a potentially mispredicted one, the noise localization and cleansing components in \tech{} are launched to improve the model performance on-the-fly.
Specifically, we leverage {\em randomized smoothing}~\cite{cohen2019certified}, which is widely studied in the field of Computer Vision that aims to certify the classification result of a given input by checking the results of a large number of random samples in the neighborhood~\cite{kumar2020certifying,mehra2021robust,chen2022input}.
In our context, we determine whether a classification is mispredicted by checking the consistency 
of the prediction results on a set of randomly perturbed samples around a given code snippet and the prediction result on the given code snippet.
We conducted an extensive study to evaluate \tech{} based on three popular pre-trained code models (i.e., CodeBERT~\cite{feng2020codebert}, GraphCodeBERT~\cite{guo2021graphcodebert}, and CodeT5~\cite{wang2021codet5}) and six code-based datasets with great diversity (involving different software engineering tasks, different scales, and different programming languages).
In total, we used 18 deep code models by fine-tuning each pre-trained model on each dataset as subjects.
Our results demonstrate the effectiveness of \tech{}.
Specifically, \tech{} corrects 21.91\% misprediction cases on average across all the subjects.
For sufficient evaluation, we still compared \tech{} with the widely-used fine-tuning strategy, even though the latter is not applicable to on-the-fly improving model performance in deployment.
The latter corrects only 9.55\% misprediction cases on average.
Besides, efficiency is also critical for on-the-fly improving model performance.
On average, \tech{} takes only 0.48 second to denoise an input, demonstrating its practicality for deployed models.


To sum up, our work makes four major contributions:
\begin{itemize}
    \item \textbf{Idea}: We propose a novel perspective to on-the-fly improve the performance of deep code models through input denoising, which is efficient, explicable, and applicable to deployed code models.

    \item \textbf{Approach}: We propose the first technique \tech{} realizing this idea, which consists of estimating a mispredicted input, localizing noise in the input code, and cleansing noise to correct the misprediction.

    \item\textbf{Evaluation}: We conducted an extensive study on 18 deep code models, demonstrating the effectiveness and efficiency of \tech{} to on-the-fly improve the performance of (deployed) deep code models.

    \item \textbf{Artifact}: We released all the experimental data and our source code at the project homepage~\cite{codedenoise2023} for experiment replication, future research, and practical use.
    
\end{itemize}

\section{Background and Motivation}
\label{sec:background}

\subsection{Deep Code Models}
\label{sec:deep_code_models}
DL has been widely used to process source code and has achieved remarkable success~\cite{wei2017supervised,li2018code,huang2018api,zhang2019novel,tian2022learning}.
Over the years, some advanced pre-trained code models have been built based on large code corpora, among which CodeBERT~\cite{feng2020codebert}, GraphCodeBERT~\cite{guo2021graphcodebert}, and CodeT5~\cite{wang2021codet5} are often treated as the representatives~\cite{jha2022codeattack,wang2023one,saberi2023model}.
CodeBERT is a large bimodal model for programming languages and natural languages, GraphCodeBERT is a structure-aware model based on semantic structure and data flow information of code, and CodeT5 is a unified encoder-decoder model that learns code semantics via an identifier-aware pre-training objective. 
The current state-of-the-art models are mainly designed based on the Transformer architecture~\cite{vaswani2017attention}, which adopts the mechanism of self-attention with the aim of differentially weighting the contribution of each part in the input (i.e., each code element in our task)~\cite{bahdanau2014neural}.
In general, these pre-trained models have been successfully applied to solve both classification tasks (such as functionality classification~\cite{zhang2019novel}) and generation tasks (such as code completion~\cite{li2018code}).
Following the recent studies on testing deep code models~\cite{henke2022semantic,yang2022natural,zhang2022towards}, our work also focuses on classification tasks and takes generation tasks as the future work. 

\subsection{Problem Definition}
\label{sec:problem_definition}
Given a code snippet $x$, a deep code model $\mathcal{M}$ can predict a probability vector, each element in which represents the probability classifying $x$ to the corresponding class.
The class with the largest probability is the final prediction result of $\mathcal{M}$ for $x$, denoted as $\mathcal{M}(x)$.
If $\mathcal{M}(x)$ is different from the ground-truth label (denoted as $c$) of $x$, it means that $\mathcal{M}$ makes a misprediction on $x$; otherwise, $\mathcal{M}$ makes a correct prediction.


Although the deep code model have achieved great performance on the training and validation sets, it cannot guarantee the correct prediction for each test input in deployment.
Our work aims to on-the-fly improve the model performance on test inputs through input denoising, which makes it an online technique different from others that require offline retraining/fine-tuning~\cite{henke2022semantic,zhang2022towards}. 
During practical use of a deep code model, we cannot exactly determine whether the prediction for a given code snippet is correct or not without manually analyzing the input and then deciding the ground-truth label.
That means it is difficult to automatically decides if
a given code snippet should be denoised.
Hence, in our work, we should first estimate whether the given code snippet $x$ is mispredicted by $\mathcal{M}$, i.e., $\mathcal{M}(x) \neq c$.
If we estimate that $x$ is correctly predicted, $\mathcal{M}(x)$ is directly outputted;
Otherwise, we start the input denoising process to correct the misprediction of $\mathcal{M}$ on $x$.
That is, we aim to obtain an input $x'$ satisfying $\mathcal{M}(x') = c \wedge x' \in \epsilon$, where $\epsilon$ is the universal set of code snippets that satisfy the grammar constraints and preserve the semantics of $x$.
$\mathcal{M}(x')$, instead of $\mathcal{M}(x)$, is the final output of $\mathcal{M}$ on $x$.
In this way, the model performance can be improved on-the-fly without retraining/fine-tuning the deep code model $\mathcal{M}$.

Here, we explain the necessity of estimating mispredicted inputs for denoising, rather than denoising each test input.
This is because denoising the inputs that are correctly predicted (1) can incur unnecessary costs, and (2) may incur risk that makes the original correct prediction become misprediction.



\subsection{A Motivating Example}
\label{sec:motivating_example}

\begin{figure}[t]
  \centering
    \subfigure[Noisy Code]{\includegraphics[width=0.233\textwidth]{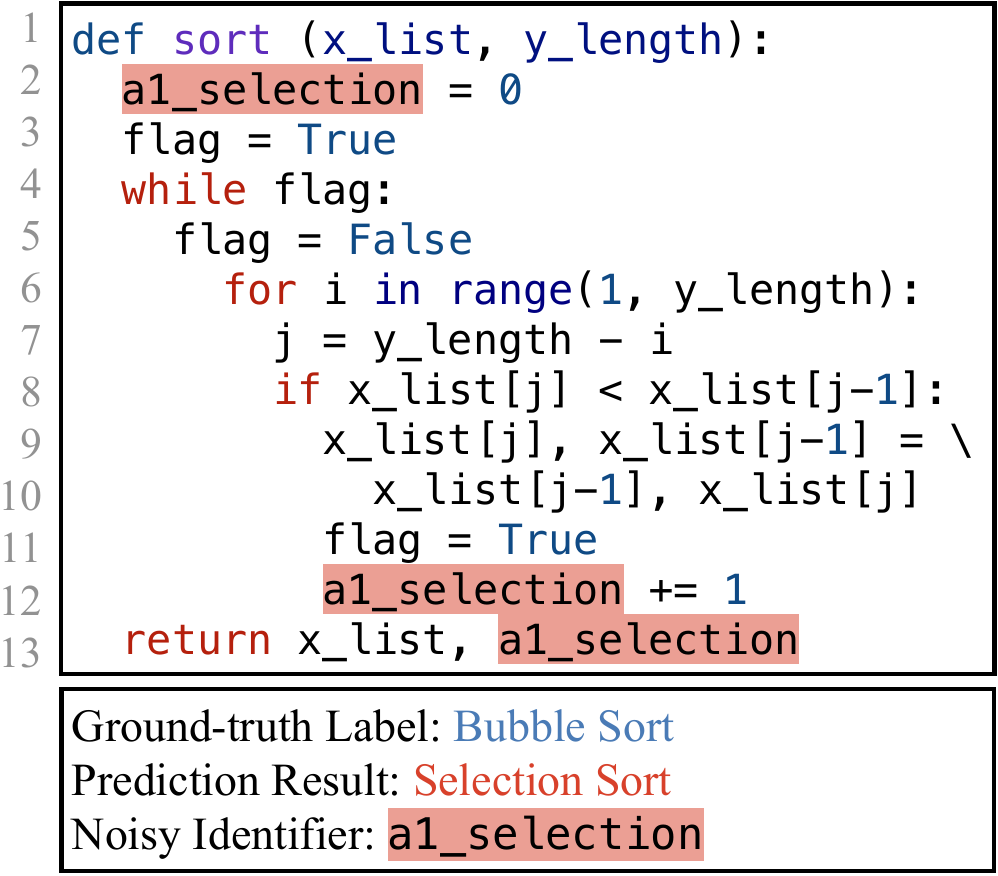}
    \label{fig:motivation_1}} 
    \subfigure[Denoised Code]{\includegraphics[width=0.233\textwidth]{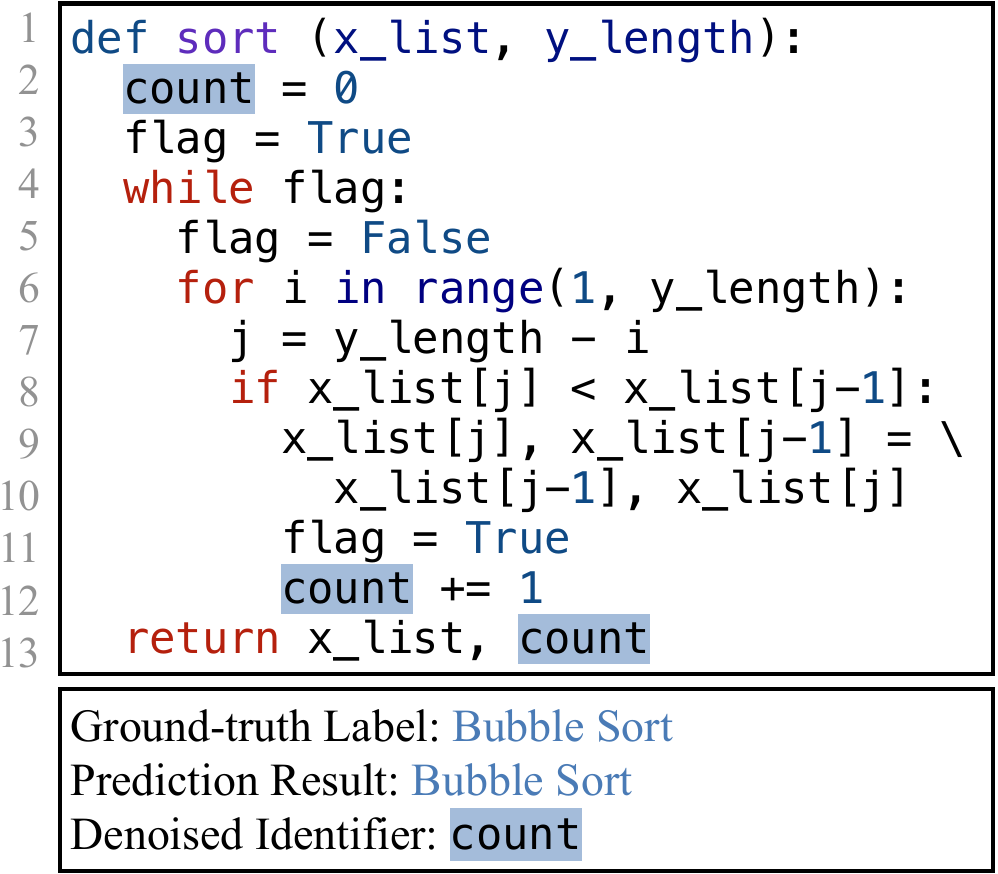}
    \label{fig:motivation_2}}
  \caption{A simplified example
  }
  \label{fig:rq1}
\end{figure}

We use a simplified example to motivate the key idea of input denoising for on-the-fly improving performance of deep code models, including noise localization (Section~\ref{sec:noise_localization}) and noise cleansing (Section~\ref{sec:noise_cleansing}) in \tech{}.
We have explained the necessity of identifying mispredicted inputs in \tech{} (Section~\ref{sec:mispredicted_input_identification}).

Given that a deep code model for the task of functionality classification, which is fine-tuned based on the pre-trained CodeBERT~\cite{feng2020codebert} with the Python800 dataset~\cite{puri2021codenet}, is deployed for practical use, there is a code snippet mispredicted by the model as shown in Figure~\ref{fig:motivation_1}.
The deployed model mispredicts the functionality of the code snippet as \textit{Selection Sort}, while its ground-truth functionality is \textit{Bubble Sort}.
Through the component of noise localization in \tech{}, we obtained that the identifier {\tt a1}\_{\tt selection} makes the largest contribution to the misprediction (which is highly relevant to the misprediction of \textit{Selection Sort} in semantics).
By replacing the noisy identifier {\tt a1}\_{\tt selection} with a clean identifier predicted by the noise cleansing component in \tech{} (i.e., {\tt count}), a denoised code snippt is obtained as shown in Figure~\ref{fig:motivation_2}.
Indeed, the deep code model corrects the misprediction based on the denoised code snippet.
This motivates \textit{the potential of on-the-fly improving performance of (deployed) deep code models through input denoising}.
In particular, by comparing the noisy identifier ({\tt a1}\_{\tt selection}) and the clean one ({\tt count}), it is more explicable for the misprediction and the correction.

Besides {\tt a1}\_{\tt selection}, the mispredicted code snippet contains the other five identifiers.
We treated each identifier as the noisy one in turn, and applied the noise cleansing component in \tech{} to predict a clean identifier to replace it.
We hence obtained five additional denoised code snippets.
However, we found that all of them cannot make the deep code model produce the correct prediction.
That motivates \textit{the necessity of identifying noisy identifiers}.
If just randomly sampling an identifier for input denoising, the worst case is to enumerate all the six identifiers to find the real noise (i.e., {\tt a1}\_{\tt selection}), which will largely slow down the process. 

Although the noisy identifier is found (through the noise localization component in \tech{}), it is still challenging to find a proper identifier to cleanse the noise within the enormous identifier space.
We randomly selected 20 identifiers from the whole set of identifiers in the Python800 dataset to replace the noisy identifier, respectively.
We found none of them correct the misprediction, and even introduce new noise making the original misprediction become other mispredicted results.
This motivates \textit{the necessity of finding an effective method to cleanse the localized noise}, which can significantly improve the input denoising effectiveness and efficiency.


\section{Approach}
\label{sec:approach}
\begin{figure}[t]
    \centering
    \includegraphics[width=1.0\linewidth]{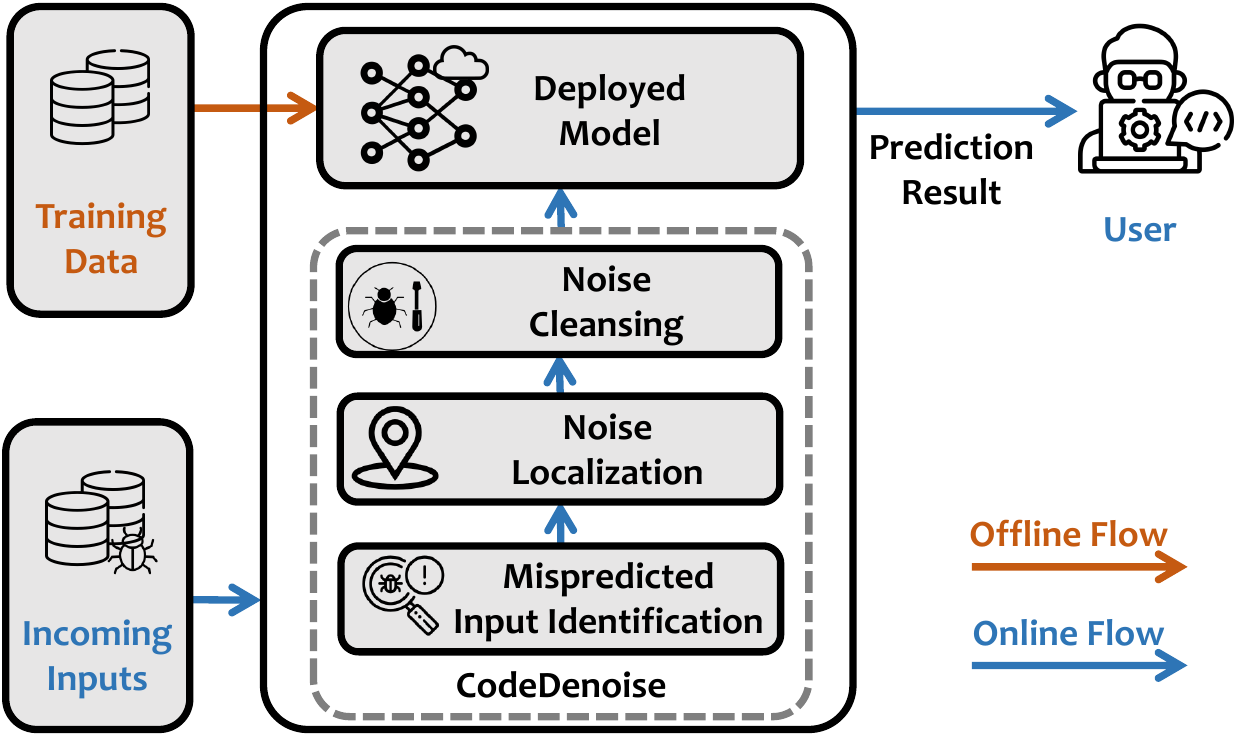}
    \caption{Overview of \tech{}}
    \label{fig:overview}
\end{figure}

\subsection{Overview}
\label{sec:overview}
Instead of retraining or fine-tuning the deep code model, in this work, we propose the first input denoising technique (called \textbf{\tech{}}) to on-the-fly improve (deployed) model performance.
Figure~\ref{fig:overview} shows the overview of \tech{}.
It consists of three main components:
mispredicted input identification (Section~\ref{sec:mispredicted_input_identification}), noise localization (Section~\ref{sec:noise_localization}), and noise cleansing (Section~\ref{sec:noise_cleansing}).
We expect that \tech{} can correct as many mispredictions as possible, but damage as few correct predictions as possible.

Here, we illustrate the usage of \tech{} in practice.
We treat \tech{} as a post-processing module and integrate it with the (deployed) code model as a system for making predictions in practice.
The system receives an incoming code snippet and outputs the prediction result.
If this code snippet is identified as a potentially mispredicted one for the deep code model by \tech{}, the system output is the prediction result corrected by \tech{};
Otherwise, the output is the original prediction result of the model.
In particular, \tech{} just accesses the deep code model but not modifies it, and thus it can accomplish the goal of on-the-fly improving performance of (deployed) models.

\subsection{Mispredicted Input Identification}
\label{sec:mispredicted_input_identification}
For an incoming code snippet, \tech{} estimates whether it is mispredicted following the idea of randomized smoothing~\cite{cohen2019certified}.
Randomized smoothing is originally proposed for image processing models.
Its key assumption is that only if most of the prediction results on a set of randomly perturbed images around a given image are the same as the prediction result on the given image, it treats the prediction made by the model reliable for the given image~\cite{cohen2019certified};
Otherwise, the prediction is unreliable.
In particular, it uses a certified radius $\mathcal{R}$ to limit the degree of perturbation on the given image.
Our idea is hence to determine if a prediction is likely wrong by checking if its perturbed versions yield the same prediction.

The original randomized smoothing method is designed specific to continuous image space, but the inputs of deep code models are source code, which actually form discrete input space.
Moreover, source code has to strictly stick to complex syntactic and semantics constraints.
Therefore, we have to re-design the randomized smoothing process specific to our task by addressing two adaptation problems:

(1) \textit{How to define the perturbation strategy on input code in \tech{}?}
The existing studies have demonstrated that identifiers have large influence on deep code models~\cite{zhang2020generating,yang2022natural,zhang2022towards}.
Moreover, identifiers are one of the most fine-grained code elements, facilitating slight perturbations.
Therefore, \tech{} perturbs input code by renaming identifiers.
We denoted the set of candidate identifiers in the vocabulary of $\mathcal{M}$ as $\mathcal{V}$ and the whole set of identifiers in the given code snippet $x$ as $\mathcal{S}(x)$.
When \tech{} renames an identifier in a $x$, denoted as $s$ ($s \in \mathcal{S}(x)$), it selects an identifier from $\mathcal{T}(x,s)$ = $\mathcal{V}$ - $\mathcal{S}(x)$ to replace $s$.
It can ensure that the identifier from $\mathcal{T}(x,s)$ does not exist in $x$, which can guarantee the syntactic correctness of the code snippet after identifier renaming and preserve the semantics of $x$.

(2) \textit{How to control the perturbation degree on input code?}
The original randomized smoothing limits the Euclidean distance between the original image and the perturbed image within $\mathcal{R}$.
That is, the perturbed images are within a distance $\leq$ $\mathcal{R}$~\cite{cohen2019certified}.
However, for discrete input space of deep code models, it may be infeasible to define such a distance like in continuous image space.
To adapt randomized smoothing to our task, \tech{} uses the number of renamed identifiers (denoted as $\delta$) to limit the perturbation degree for a given code snippet.
That is, $\delta$ should be smaller than a threshold $\theta$ (that plays the same role as the radius $\mathcal{R}$).

\tech{} adopts the above adapted solutions to estimate whether an input code $x$ is mispredicted.
Specifically, it constructs $N$ perturbed code snippets for $x$, each of which randomly renames $\delta$ identifiers in $x$ ($\delta$ is a random positive number satisfying $\delta \leq \theta \wedge \delta \leq |\mathcal{S}(x)|$).
By comparing the prediction results of $x$ and the $N$ perturbed code snippets,
if the majority prediction result of the $N$ perturbed code snippets is not the same as $\mathcal{M}(x)$, \tech{} identifies $x$ a mispredicted input and then starts the follow-up components to denoise $x$.
There are two important hyper-parameters $\theta$ and $N$, and we will investigate their influence in Section~\ref{sec:rq3}.

\subsection{Noise Localization }
\label{sec:noise_localization}
When identifying a code snippet as a mispredicted one, \tech{} then localizes noise in it to facilitate the follow-up cleansing process.
As explained aforementioned, \tech{} focuses on identifier-level noise as the first exploration in this direction.
There may be other levels of noise in code (such as code structure noise), and we take them as the future work.
Hence, the noise in our work refers to identifiers contributing to mispredictions (called {\em noisy identifiers}).
Cleansing noisy identifiers can help correct mispredictions and thus improve model performance.


\tech{} relies on the built-in attention mechanism in deep code models to localize noisy identifies in the input code.
Although the attention mechanism is not a must for all models, almost all state-of-the-art models use it (in particular, it is the core of the state-of-the-art Transformer architecture~\cite{vaswani2017attention}), since (1) it is helpful to improve model performance and (2) it provides an opportunity to explain model prediction.
Specifically, it assigns a weight to each part of an input (i.e., each token in our task), indicating the contribution of the part to the prediction. 

In fact, state-of-the-art models often contain a number of attention layers and the attention weights are calculated at each attention layer. 
That is, the contribution of each token in the input code is measured at each attention layer. 
For ease of presentation, we denote the list of tokens in the mispredicted code snippet as $\mathcal{E} = \{e_1, e_2, \ldots, e_m\}$, the list of attention layers as $\mathcal{H} = \{h_1, h_2, \ldots, h_n\}$. 
The attention weight of $e_i$ at the attention layer $h_j$ is denoted as $w_{ij}$ 
(more details on attention weight calculation can be found in~\cite{vaswani2017attention}).
Following the existing work~\cite{zhang2022diet}, \tech{} calculates the contribution of each token by aggregating the corresponding attention weights at all the layers as $W_i=\frac{\sum_{j=1}^{n}{w_{ij}}}{n}$.


Since \tech{} just considers identifier-level noise, it discards tokens that are not identifiers (such as keywords and operators).
Besides, different tokens may represent the same identifier, e.g., {\tt a1}\_{\tt selection} appears in three token positions in the code snippet in Figure~\ref{fig:motivation_1}, and thus \tech{} further aggregates token-level contributions to identifier-level contributions.
Here, \tech{} uses the largest contribution among the tokens corresponding to an identifier as the contribution of the identifier.
Then, \tech{} ranks all the identifiers in the descending order of their contributions, indicating that the identifiers with larger contributions to the misprediction are more likely to be noise in the code snippet.

\begin{algorithm}[t]
	\caption{Noise Cleansing}
	\label{alg:algorithm2}
    \KwIn{
    \\ \qquad $x$: identified mispredicted code;
    \\ \qquad $c$: original prediction result;
    \\ \qquad $\mathcal{M}$: target model;
    \\ \qquad $\mathcal{M}_{\textit{mcip}}$: MCIP model;
    \\ \qquad $\mathcal{S}(x)$: the set of identifiers of $x$;
    \\ \qquad $\mathcal{V}$: the set of candidate identifiers; 
    \\ \qquad $idenList$: the ranking list of noisy identifiers;
    }

    \KwOut{ \\ \qquad $c'$: the prediction result after denoising}
    
	\BlankLine
    $c' = c$; \\
    {\small \tcp{get model's confidence of $x$ on $c$}}
    $prob$ = {\tt probability}($x$, $\mathcal{M}$)[$c$]; \\
    \ForEach{$iden$ in $idenList$}{
        $iden'$ = $\mathcal{M}_{mcip}$({\tt replace}($x$, $iden$, {\tt <mask>})); \\
        \If{$iden' \notin \mathcal{S}(x)$}{
            {\small \tcp{replace identifiers for input code}}
            $x'$ = {\tt replace}($x$, $iden$, $iden'$); \\
            $c' = \mathcal{M}(x')$; \\
            \If{$c' \neq c$}{
                \textbf{return} $c'$;
            }
            $prob'$ = {\tt probability}($x'$, $\mathcal{M}$)[$c$]; \\
            \If{$prob' < prob$}{
                $prob = prob'$; \\
                $x = x'$; \\
            }
        }
    }
	\textbf{return} $c'$;
\end{algorithm}

\subsection{Noise Cleansing}
\label{sec:noise_cleansing}
After localizing noisy identifiers, \tech{} further cleanses noise to correct the misprediction.
The basic idea of noise cleansing in \tech{} is to replace noisy identifiers with clean identifiers.
It can be achieved by first masking noisy identifiers in the mispredicted code snippet and then generating clean identifiers at the masked locations.
This process follows the same workflow as the task of masked identifier prediction (MIP)~\cite{feng2020codebert,jiang2023impact}, which also aims to predict the tokens at the masked locations.
However, the existing MIP models are not aligned with our goal.
Specifically, they predict an identifier \textit{natural} to the context at a masked location in the code snippet, but do not consider whether the identifier is \textit{clean}.
In fact, the training data of the existing MIP models may also include noise, which may result in predicting natural but noisy identifiers.
Therefore, they cannot achieve the goal of correcting misprediction, even can incur new noise to aggravate misprediction.

To predict a clean identifier to replace the noisy identifier, \tech{} builds a \textit{masked clean identifier prediction} (MCIP) model on the basis of some MIP model.
Specifically, \tech{} fine-tunes the MIP model based on clean data with regard to the deep code model.
Here, \tech{} removes the data that are mispredicted by the deep code model from the training data corresponding to the model, and then regards the remaining data as clean data for fine-tuning the MIP model.
In this way, a MCIP model is built, which helps cleanse noisy identifiers in the mispredicted code snippet.
In \tech{}, we choose CodeBERT~\cite{feng2020codebert} as the underlying MIP model.
However, \tech{} is not specific to CodeBERT and we will evaluate \tech{} under different MIP models in the future.
Note that, for each deep code model, \tech{} builds a MCIP model by fine-tuning the underlying MIP model, but the model is built offline and can be deployed together with the target deep code model.


Based on the ranking list of noisy identifiers from the noise localization component and the MCIP model, \tech{} starts the noise cleansing process for the mispredicted code snippet.
This process is iterative as illustrated in Algorithm~\ref{alg:algorithm2}.
At each iteration, \tech{} determines a noisy identifier according to the ranking list and then masks it to feed the MCIP model (Line 5).
The model then predicts an identifier to fill this masked location to form a new code snippet (Line 8).
It is possible for the model to predict an identifier that has already existed in the mispredicted code snippet, which can introduce syntactic errors.
Hence, \tech{} discards the new code snippet and jumps to the next iteration (Line 6).

If the new code snippet is valid, \tech{} feeds it to the deep code model (Line 9).
As long as the new code snippet changes the original prediction result, \tech{} terminates the cleansing process and outputs the prediction result of the new code snippet, which is expected to correct the original misprediction via input denoising (Lines 10-12).
Although it does not change the original prediction result, if it makes the confidence of the original predicted result decrease, \tech{} puts the new code snippet to the next iteration and the follow-up cleansing process is conducted on it (instead of the original code snippet) (Lines 13-17).
Otherwise, \tech{} discards the new code snippet and jumps to the next iteration.
This is a greedy process.
In the future, more advanced search algorithms can be incorporated to improve the cleansing process.
Besides finding the code snippet that changes the original prediction, \tech{} terminates after all the noisy identifiers are enumerated for denoising via the MCIP model, and still outputs the original prediction result.
Note that \tech{} replaces all occurrences of a noisy identifier in the code with the clean identifier, which ensures that the code semantics is not changed.

\section{Evaluation Design}
\label{sec:design}
Our study aims to address the following research questions:

\begin{itemize}
    \item \textbf{RQ1}: Can \tech{} improve the model performance on-the-fly effectively and efficiently?
 
    \item \textbf{RQ2}: Does each main component in \tech{} contribute to its overall effectiveness?
    
    \item \textbf{RQ3}: How does \tech{} perform under different hyper-parameters' configurations?
\end{itemize}

\begin{table}[t]
\caption{Statistics of our used subjects}
\label{tab:tasks_and_models}
\centering
\tabcolsep=0.8mm
\begin{adjustbox}{max width=0.5\textwidth,center}
    \begin{threeparttable}
        \begin{tabular}{ ccrrrr }
        \toprule
        \textbf{Task} & \textbf{Train/Validate/Test} & \textbf{Class} & \textbf{Language} & \textbf{Model} & \textbf{Acc.}  \\ \midrule
        \multirow{3}{*}{\makecell[c]{Authorship \\ Attribution}} & \multirow{3}{*}{ 528/ - /132 } & \multirow{3}{*}{66} & \multirow{3}{*}{Python} & CB & 82.58\% \\
         &&&& GCB & 77.27\% \\
         &&&& CT5 & 83.33\% \\
        \hline
        \multirow{3}{*}{\makecell[c]{Defect \\ Prediction}} & \multirow{3}{*}{ 27,058/ - /6,764 } & \multirow{3}{*}{4} & \multirow{3}{*}{C/C++} & CB & 85.47\% \\
         &&&& GCB & 83.90\% \\
         &&&& CT5 & 82.29\% \\
        \hline
        \multirow{3}{*}{\makecell[c]{Functionality \\ Classification \\ C104}} & \multirow{3}{*}{ 41,581/ - /10,395 } & \multirow{3}{*}{104} & \multirow{3}{*}{C} & CB & 97.87\% \\
         &&&& GCB & 98.61\% \\
         &&&& CT5 & 98.60\% \\
        \hline
        \multirow{3}{*}{\makecell[c]{Functionality \\ Classification \\C++1000}} & \multirow{3}{*}{ 320,000/80,000/100,000 } & \multirow{3}{*}{1,000} & \multirow{3}{*}{C++} & CB & 85.00\% \\
         &&&& GCB & 81.62\% \\
         &&&& CT5 & 86.49\% \\
        \hline
        \multirow{3}{*}{\makecell[c]{Functionality \\ Classification \\Python800}} & \multirow{3}{*}{ 153,600/38,400/48,000 } & \multirow{3}{*}{800} & \multirow{3}{*}{Python} & CB & 93.91\% \\
         &&&& GCB & 97.39\% \\
         &&&& CT5 & 97.62\% \\
        \hline
        \multirow{3}{*}{\makecell[c]{Functionality \\ Classification \\Java250}} & \multirow{3}{*}{ 48,000/11,909/15,000 } & \multirow{3}{*}{250} & \multirow{3}{*}{Java} & CB & 96.30\% \\
         &&&& GCB & 97.79\% \\
         &&&& CT5 & 97.48\% \\
        \bottomrule
        \end{tabular}
        \begin{tablenotes}
            \footnotesize
            \item[*] CB/GCB/CT5 is short for CodeBERT/GraphCodeBERT/CodeT5.
        \end{tablenotes}
    \end{threeparttable}
\end{adjustbox}
\end{table}

\subsection{Subjects}
\label{sec:sub}
To sufficiently evaluate \tech{}, we used three state-of-the-art pre-trained models (i.e., CodeBERT~\cite{feng2020codebert}, GraphCodeBERT~\cite{guo2021graphcodebert}, and CodeT5~\cite{wang2021codet5}) and six code-based datasets (i.e., Authorship Attribution~\cite{alsulami2017source}, Defect Prediction~\cite{codechef2022}, Functionality Classification C104~\cite{zhang2019novel}, Functionality Classification C++1000~\cite{puri2021codenet}, Functionality Classification Python800~\cite{puri2021codenet}, and Functionality Classification Java250~\cite{puri2021codenet}) in our study.
These models and datasets have been widely used in many existing studies on evaluating the robustness of deep code models~\cite{yang2022natural,zhang2022towards,zhang2023challenging}.
By fine-tuning each pre-trained model on each dataset, we obtained 18 deep code models as subjects in total.
Table~\ref{tab:tasks_and_models} shows the statistics of our subjects, each column in which represents the dataset (indicating the task), the number of inputs in the training/validation/test set, the number of classes for the classification task, the programming language for the inputs, the used pre-trained model, and the accuracy of the deep code model after fine-tuning the corresponding pre-trained model on the corresponding dataset, respectively.
The detailed hyper-parameter settings of these deep code models can be found at our project homepage~\cite{codedenoise2023}.

\textit{Authorship Attribution} aims to identify the author of a given code snippet. 
Its used dataset is the Google Code Jam (GCJ) dataset~\cite{alsulami2017source}.
\textit{Defect Prediction} aims to predict whether a given code snippet is defective and its defect type.
Its used dataset is the CodeChef dataset~\cite{codechef2022}, which is labeled by the execution results on the CodeChef platform.
\textit{Functionality Classification} aims to classify the functionality of a given code snippet.
If code snippets solve the same problem, they are regarded to have the same functionality~\cite{zhang2019novel}. 
Regarding the task of functionality classification, we used four different datasets.
\textit{C104} is the Open Judge (OJ) benchmark~\cite{mou2016convolutional}, which involves 104 classes of C code snippets and has been integrated as part of the CodeXGLUE benchmark~\cite{lu2021codexglue}.
\textit{C++1000}, \textit{Python800}, and \textit{Java250} are more large-scale and provided by the CodeNet project~\cite{puri2021codenet}, which involve 1,000 classes of C++ code snippets, 800 classes of Python code snippets, and 250 classes of Java code snippets, respectively.
These datasets contain various lengths of code snippets (ranging from 6 to 4,062 tokens with an average of 271 tokens across the six datasets).



Overall, the subjects used in our study are diverse, which involve different tasks, different pre-trained models, different numbers of classes, different programming languages, and different scales.
It is quite helpful to sufficiently evaluate \tech{} and thus demonstrate the generality of \tech{}.

\subsection{Implementations}
\label{sec:implementations}

We implemented \tech{} in Python and adopted tree-sitter~\cite{treesitter2022} to extract identifiers from code snippets.
By default, we set $\theta$ (the threshold of the number of renamed identifiers to limit the perturbation degree) to 1 and $N$ (the number of perturbed code snippets) to the number of identifiers in the given code snippet in \tech{} by conducting a preliminary experiment.
We also investigated the influence of the settings for the two hyper-parameters in Section~\ref{sec:rq3}.
\textit{To reduce the influence of randomness, we repeated all the experiments 5 times, and reported the average results.}
All the experiments were conducted on an Ubuntu 20.04 server with Intel(R) Xeon(R) Silver 4214 @ 2.2GHz CPU, 256GB of RAM, and NVIDIA GeForce RTX 2080 Ti GPU.

\subsection{Measurements}
\label{sec:metric}
Our work aims to improve performance of a deep code model by trying to correct mispredictions through input denoising.
Hence, we measured the effectiveness of \tech{} in terms of both \textbf{correction success rate} (\textbf{CSR}) and \textbf{mis-correction rate} (\textbf{MCR}).
CSR measures the ability of \tech{} to correct mispredictions, which is the ratio of the number of inputs whose mispredictions are successfully corrected by \tech{} to the total number of mispredicted inputs in the test set.
It is possible that the mispredicted inputs identified by \tech{} are actually predicted correctly.
Denoising such inputs may incur the risk that makes the original correct prediction become misprediction.
We call these cases \textit{mis-corrections}.
MCR measures the negative effect caused by \tech{}, which is the ratio of the number of inputs whose correct predictions are changed to mispredictions by \tech{} to the total number of inputs with correct predictions in the test set.
Larger CSR values and smaller MCR values mean better effectiveness.
In particular, we also measured the \textbf{overall accuracy} of the deep code model after input denoising to show the model performance improvement brought by \tech{} in a more intuitive manner.



Besides, we used \textbf{the time spent on input denoising} and \textbf{the number of changed identifiers} to measure the efficiency of \tech{} in improving model performance.
Less time and fewer identifier changes mean higher efficiency.
High efficiency is important for on-the-fly improving performance of the (deployed) model; otherwise the user experience for the deep code model can be negatively affected.
In particular, the number of changed identifiers is not only related to the number of noise-cleansing iterations, but also related to the naturalness of denoised inputs. 
Fewer changes tend to mean more natural denoised inputs~\cite{yang2022natural}.
\begin{table*}[t]
    \caption{Effectiveness comparison in terms of CSR ($\uparrow$)/MCR ($\downarrow$)}
    \label{tab:csr_mcr}
    \centering
    \tabcolsep=2.3mm
    \begin{adjustbox}{max width=1.0 \textwidth,center}
        \begin{tabular}{ ccccccc }
            \toprule
        	\multirow{2}{*}{\textbf{Task}} & \multicolumn{2}{c}{\textbf{CodeBERT}} & \multicolumn{2}{c}{\textbf{GraphCodeBERT}} & \multicolumn{2}{c}{\textbf{CodeT5}} \\ \cmidrule(lr){2-3} \cmidrule(lr){4-5} \cmidrule(lr){6-7}
        	& \textbf{Fine-tuning} & \textbf{\tech{}} & \textbf{Fine-tuning} & \textbf{\tech{}} & \textbf{Fine-tuning} & \textbf{\tech{}} \\
        	\midrule
                Authorship Attribution & 20.00\%/1.79\% & \textbf{30.00\%}/\textbf{0.00\%} & 10.00\%/0.00\% & \textbf{20.00\%}/\textbf{0.00\%} & 10.00\%/1.79\% & \textbf{40.00\%}/\textbf{0.00\%} \\
                Defect Prediction & \enspace 5.98\%/0.59\% & \textbf{22.47\%}/\textbf{0.24\%} & \enspace 8.51\%/1.44\% & \textbf{28.73\%}/\textbf{0.18\%} & \enspace 5.15\%/0.29\% & \textbf{16.64\%}/\textbf{0.18\%} \\
                Functionality Classification C104& \enspace 7.32\%/0.08\% & \textbf{17.07\%}/\textbf{0.02\%} & \enspace 5.88\%/0.06\% & \textbf{14.12\%}/\textbf{0.04\%} & 13.41\%/0.06\% & \textbf{15.85\%}/\textbf{0.04\%} \\
                Functionality Classification C++1000 & \enspace 1.42\%/0.17\% & \textbf{27.32\%}/\textbf{0.05\%} & \enspace 1.95\%/0.34\% & \textbf{\enspace 5.14\%}/\textbf{0.05\%} & \enspace 1.14\%/0.15\% & \textbf{14.13\%}/\textbf{0.04\%} \\
                Functionality Classification Python800 & \enspace 4.19\%/0.09\% & \textbf{28.76\%}/\textbf{0.03\%} & \enspace 7.18\%/0.08\% & \textbf{20.48\%}/\textbf{0.05\%} & \enspace 3.35\%/0.06\% & \textbf{19.55\%}/\textbf{0.02\%} \\
                Functionality Classification Java250 & 23.00\%/0.07\% & \textbf{31.71\%}/\textbf{0.04\%} & 16.67\%/0.26\% & \textbf{23.21\%}/\textbf{0.25\%} & 26.83\%/0.03\% & \textbf{27.80\%}/\textbf{0.03\%} \\
        	\midrule
        	Average & 10.32\%/0.46\% & \textbf{26.22\%}/\textbf{0.06\%} & \enspace 8.37\%/0.36\% & \textbf{18.61\%}/\textbf{0.09\%} & \enspace 9.98\%/0.39\% & \textbf{22.33\%}/\textbf{0.05\%} \\
            \bottomrule
        \end{tabular}
    \end{adjustbox}
\end{table*}

\begin{table*}[t]
    \caption{Effectiveness comparison in terms of the overall accuracy ($\uparrow$) of model (Ori: the accuracy of the original models)}
    \label{tab:accuracy}
    \centering
    \tabcolsep=1.0mm
    \begin{adjustbox}{max width=1.0 \textwidth,center}
        \begin{tabular}{ cccccccccc }
            \toprule
        	\multirow{2}{*}{\textbf{Task}} & \multicolumn{3}{c}{\textbf{CodeBERT}} & \multicolumn{3}{c}{\textbf{GraphCodeBERT}} & \multicolumn{3}{c}{\textbf{CodeT5}} \\ \cmidrule(lr){2-4} \cmidrule(lr){5-7} \cmidrule(lr){8-10}
        	& \textbf{Ori} & \textbf{Fine-tuning} & \textbf{\tech{}} & \textbf{Ori} & \textbf{Fine-tuning} & \textbf{\tech{}} & \textbf{Ori} & \textbf{Fine-tuning} & \textbf{\tech{}} \\
        	\midrule
                Authorship Attribution & 84.85\% & 86.36\% & \textbf{89.39\%} & 84.85\% & 86.36\% & \textbf{87.88\%} & 84.85\% & 84.85\% & \textbf{90.91\%} \\
                Defect Prediction & 85.66\% & 86.01\% & \textbf{88.68\%} & 84.36\% & 84.48\% & \textbf{88.70\%} & 82.76\% & 83.41\% & \textbf{85.48\%} \\
                Functionality Classification C104 & 97.63\% & 97.73\% & \textbf{98.02\%} & 98.36\% & 98.40\% & \textbf{98.56\%} & 98.42\% & 98.58\% & \textbf{98.63\%} \\
                Functionality Classification C++1000 & 84.93\% & 85.00\% & \textbf{89.00\%} & 81.68\% & 81.77\% & \textbf{82.59\%} & 86.50\% & 86.52\% & \textbf{88.37\%} \\
                Functionality Classification Python800 & 97.12\% & 97.15\% & \textbf{97.92\%} & 98.43\% & 98.46\% & \textbf{98.71\%} & 97.76\% & 97.78\% & \textbf{98.18\%} \\
                Functionality Classification Java250 & 96.17\% & 96.99\% & \textbf{97.35\%} & 97.76\% & 97.88\% & \textbf{98.04\%} & 97.27\% & 97.97\% & \textbf{98.00\%} \\
        	\midrule
        	Average & 91.06\% & 91.54\% & \textbf{93.39\%} & 90.91\% & 91.23\% & \textbf{92.42\%} & 91.26\% & 91.52\% & \textbf{93.26\%} \\
            \bottomrule
        \end{tabular}
    \end{adjustbox}
\end{table*}

\section{Results and Analysis}
\label{sec:results}

 \subsection{RQ1: Effectiveness and Efficiency of \tech{}}
\label{sec:rq1}

\subsubsection{Baseline}
\tech{} is the first technique to on-the-fly improve the performance of deep code models, and thus we cannot find direct baselines for comparison.
To better demonstrate the effectiveness and efficiency of \tech{}, we compared \tech{} with the model fine-tuning strategy, which is the most widely-used strategy for improving the model performance but cannot achieve the goal of on-the-fly improving the performance of (deployed) models.

\subsubsection{Process}
For each subject, we divided the test set into two parts equally, denoted as $T_1$ and $T_2$.
We treated $T_1$ as the fine-tuning set while $T_2$ as the evaluation set.
For the baseline, we fine-tuned the deep code model with $T_1$ to obtained a fine-tuned model based on the same settings and process used for the original model training.
Then, we evaluated the baseline on $T_2$.
For fair comparison, we also evaluated \tech{} on $T_2$.
Finally, we compared \tech{} and the fine-tuning strategy in terms of the above-mentioned metrics.


\subsubsection{Results}
Table~\ref{tab:csr_mcr} shows the comparison results between the fine-tuning strategy and \tech{} in terms of CSR and MCR.
From this table, we found that \tech{} always performs better than the fine-tuning strategy on each subject in terms of CSR and MCR, demonstrating the stable effectiveness of \tech{}. 
On average, \tech{} successfully corrects 26.22\%, 18.61\%, and 22.33\% mispredictions on CodeBERT, GraphCodeBERT, and CodeT5, while the fine-tuning strategy just corrects 10.32\%, 8.37\%, and 9.98\% mispredictions, respectively.
Also, \tech{} makes only 0.06\%, 0.09\%, and 0.05\% mis-corrections on CodeBERT, GraphCodeBERT, and CodeT5 on average, while the latter makes 0.46\%, 0.36\%, and 0.39\% mis-corrections, respectively.
On average across all the 18 subjects, \tech{} improves the fine-tuning strategy by 330.02\% in terms of CSR and 54.12\% in terms of MCR, respectively.

Table~\ref{tab:accuracy} shows the original accuracy of each deep code model, the accuracy of each fine-tuned model, and the overall model accuracy after applying \tech{}, on $T_2$ respectively.
From this table, both the fine-tuning strategy and \tech{} improve the model accuracy on each deep code model, and meanwhile \tech{} always performs better than the fine-tuning strategy. 
Specifically, \tech{} improves 0.20\%$\sim$7.14\% higher accuracy than the original model across all the subjects, while the fine-tuning strategy just improves 0.00\%$\sim$1.78\% higher accuracy.

Overall, \tech{} is effective to improve model performance, even performs better than the state-of-the-art fine-tuning strategy.
In particular, \tech{} can on-the-fly improve performance of deep code models (in deployment), while the fine-tuning strategy is not applicable to this scenario.
In fact, both techniques are orthogonal to a large extent, which will be discussed in detail in Section~\ref{sec:discussion}.


In addition, we compared them in terms of efficiency even though the two techniques have different application scenarios.
Specifically, \tech{} takes 35.98 hours to complete the entire process for all the 18 subjects.
On average, it takes only 0.48 second to denoise an input (with the time variance of 0.05), which cannot negatively affect the practical use of the deployed model in general.
The fine-tuning strategy takes 133.22 hours to complete the entire process and it cannot be individually applied to each input.
The results demonstrate the high efficiency of \tech{} for on-the-fly improving model performance.




\begin{tcolorbox}[]
\textbf{Answer to RQ1:} \tech{} is effective and efficient to on-the-fly improve performance of (deployed) deep code models, with 22.39\% CSR and 0.07\% MCR on average by taking only 0.48 second to denoise one input.
In particular, \tech{} outperforms the fine-tuning strategy with 134.33\% larger CSR and 83.04\% smaller MCR on average.

\end{tcolorbox}

 \subsection{RQ2: Contribution of Each Main Component}
\label{sec:rq2}



\subsubsection{Variants of \tech{}}
\tech{} contains three main components: randomized-smoothing-based mispredicted input identification, attention-based noise localization, and MCIP-based noise cleansing.
To investigate the contribution of each component, we constructed a series of variants of \tech{}.

For the fist component, some existing work assumed the correlation between misprediction and classification uncertainty~\cite{feng2020deepgini,weiss2021fail,weiss2022simple}.
They assumed that the inputs whose classifications are more uncertain, are more likely to be mispredicted.
DeepGini provides a state-of-the-art method to measure classification uncertainty~\cite{feng2020deepgini}.
Given the prediction probability $p(c)$ on the class $c$ ($\in \{1, 2, ..., C\}$) for a given input, the classification uncertainty is calculated as $1-\sum _{c=1}^{C} p(c)^{2}$.
To investigate the effectiveness of randomized-smoothing-based mispredicted input identification, we replaced the randomized-smoothing-based strategy with the DeepGini-based strategy in \tech{}, which forms a variant of \tech{} called \textbf{\tech$_{\textit{deepgini}}$}.
Specifically, \tech$_{\textit{deepgini}}$ identifies a mispredicted input if its DeepGini-based uncertainty exceeds a pre-defined threshold $\zeta$.
We set $\zeta$ according to the best performance of DeepGini on the corresponding training set, i.e., selecting the threshold with the best true positive rate under $\leq$5\% false positive rate.

For the second component, we constructed a variant of \tech{} by replacing the attention-based strategy with a random strategy, called \textbf{\tech$_{\textit{randL}}$}.
Specifically, \tech$_{\textit{randL}}$ randomly selects identifiers from the mispredicted code snippet as noise.

For the third component, we constructed two variants of \tech{}, i.e., \textbf{\tech$_{\textit{randC}}$} and \textbf{\tech$_{\textit{MIP}}$}.
Instead of the MCIP-based strategy, \tech$_{\textit{randC}}$ randomly selects an identifier from the vocabulary to replace the noisy identifier for noise cleansing, while \tech$_{\textit{MIP}}$ uses the MIP model (i.e., CodeBERT) to predict an identifier to replace the noisy one for noise cleansing.


\subsubsection{Process}
Since this experiment (same as RQ3) does not involve the fine-tuning strategy, it is not necessary to divide each test set into a fine-tuning set and an evaluation set.
Therefore, we applied \tech{} as well as the four variants of \tech{} to each deep code model on the corresponding entire test set, and then measured their effectiveness and efficiency, respectively.


\subsubsection{Results}
Table~\ref{tab:ablation} shows the comparison results among all the four variants and \tech{} in terms of CSR, MCR, and the average number of changed identifiers.
By comparing \tech{} with \tech$_{\textit{deepgini}}$, \tech{} outperforms \tech$_{\textit{deepgini}}$ in terms of both CSR and MCR on almost all cases (except only one case).
For the one exceptional case, \tech{} has slightly larger MSR values than \tech$_{\textit{deepgini}}$ (0.05\% vs 0.02\%), but \tech{} has larger CSR values than the latter on this case. 
In particular, on all other cases, \tech{} improves \tech$_{\textit{deepgini}}$ by 63.14\% in terms of average MCR, 47.23\% in terms of average CSR, and 28.42\% in terms of the average number of changed identifiers for improving the model performance.
The results demonstrate that the randomized-smoothing-based strategy in \tech{} is more effective than the widely-studied DeepGini-based strategy for improving model performance.
The main reason lies in that the randomized-smoothing-based strategy identifies more really-mispredicted inputs and less correctly-predicted inputs as mispredicted inputs than the latter, and thus corrects more mispredictions and incurs less mis-corrections.
Hence, the accuracy of identifying mispredicted inputs is important for \tech{} to improve model performance.


We also applied \tech{} (without the first component) to the entire test set (taking the three deep code models with the Defect Prediction dataset as the representative) in order to investigate the necessity of identifying mispredicted inputs.
Its average MCR value is 13.32\% across the three deep code models, which is much larger than that achieved by \tech{} with the first component (only 0.28\%).
This is because \tech{} aims to cleanse the identifiers contributing to misprediction; but if \tech{} is applied to correctly-predicted inputs, the identifiers cleansed by \tech{} are actually those contributing to correct prediction.
This may incur the risk of mis-correction.
Moreover, \tech{} without the first component can be more time-consuming (5.10 hours to complete the entire process of the three models) than it with the first component (3.03 hours).
The results confirm the necessity of mispredicted input identification.

By comparing \tech{} with \tech$_{\textit{randL}}$, the former outperforms the latter on all cases in terms of the three metrics (CSR, MCR, the average number of changed identifiers), confirming the contribution of the attention-based noise localization component.
Specifically, \tech{} improves \tech$_{\textit{randL}}$ by 56.54\% in terms of average CSR, 74.61\% in terms of average MCR, and 57.87\% in terms of the average number of changed identifiers.
The main reason lies in, randomly localizing identifiers for cleansing may change the original prediction, but is less likely to change misprediction to the correct one, as the identifiers may be not real noise.

By comparing \tech{} with \tech$_{\textit{randC}}$ and \tech$_{\textit{MIP}}$, the former outperforms the latter two on all cases in terms of all the three metrics, confirming the contribution of the MCIP-based noise cleansing.
Specifically, \tech{} improves \tech$_{\textit{randC}}$ and \tech$_{\textit{MIP}}$ by 117.64\% and 43.79\% in terms of average CSR, 74.61\% and 58.57\% in terms of average MCR, and 49.50\% and 25.53\% in terms of the average number of changed identifiers, respectively.
The main reason lies in, randomly selecting identifiers from the vocabulary or applying the MIP model to predict identifiers is less likely to obtain clean identifiers for cleansing noise, even may be more likely to introduce new noise.


\begin{table*}[t]
    \caption{Ablation test for \tech{} in terms of CSR ($\uparrow$)/MCR ($\downarrow$)/the average number of identifier changes ($\downarrow$)}
    \label{tab:ablation}
    \centering
    \tabcolsep=2.3mm
    \begin{adjustbox}{max width=1.0\textwidth,center}
        \begin{tabular}{ ccccccc }
            \toprule
            \textbf{Task} & \textbf{Model} & \tech$_{\textit{deepgini}}$ & \tech$_{\textit{randL}}$ & \tech$_{\textit{randC}}$ & \tech$_{\textit{MIP}}$ & \textbf{\tech{}} \\ \midrule
            \multirow{3}{*}{\makecell[c]{Authorship \\ Attribution}} 
             & CodeBERT & 26.09\%/0.92\%/2.01 & 26.09\%/1.83\%/3.29 & 13.04\%/0.92\%/3.28 & 21.74\%/0.92\%/2.20 & \textbf{30.43\%}/\textbf{0.00\%}/\textbf{1.43} \\
             & GraphCodeBERT & 16.17\%/0.98\%/2.85 & 16.67\%/0.98\%/3.51 & 16.67\%/1.96\%/2.18 & 15.67\%/0.98\%/3.20 & \textbf{23.33\%}/\textbf{0.00\%}/\textbf{1.57} \\
             & CodeT5 & 22.73\%/1.82\%/1.67 & 31.82\%/0.91\%/3.75 & 27.27\%/0.91\%/3.73 & 31.82\%/0.91\%/1.49 & \textbf{40.91\%}/\textbf{0.00\%}/\textbf{1.44} \\
            \hline
            \multirow{3}{*}{\makecell[c]{Defect \\ Prediction}} 
             & CodeBERT & \enspace 7.32\%/0.42\%/2.21 & 14.45\%/0.43\%/3.62 & \enspace 9.26\%/0.50\%/2.35 & 13.43\%/0.48\%/2.08 & \textbf{20.45\%}/\textbf{0.35\%}/\textbf{1.35} \\
             & GraphCodeBERT & 13.59\%/1.76\%/3.17 & 19.74\%/0.51\%/4.55 & 14.97\%/0.55\%/3.02 & 22.04\%/0.48\%/3.66 & \textbf{28.37\%}/\textbf{0.30\%}/\textbf{2.02} \\
             & CodeT5 & \enspace 7.85\%/1.46\%/2.54 & \enspace 8.85\%/0.49\%/3.83 & \enspace 8.18\%/0.52\%/3.65 & \enspace 9.27\%/0.38\%/2.71 & \textbf{14.52\%}/\textbf{0.20\%}/\textbf{1.65} \\
            \hline
            \multirow{3}{*}{\makecell[c]{Functionality \\ Classification \\ C104}} 
             & CodeBERT & 19.00\%/0.23\%/1.91 & 12.22\%/0.08\%/3.22 & \enspace 9.50\%/0.07\%/3.61 & 15.38\%/0.06\%/1.75 & \textbf{19.91\%}/\textbf{0.04\%}/\textbf{1.28} \\
             & GraphCodeBERT & 15.97\%/0.11\%/2.40 & 11.81\%/0.15\%/5.69 & \enspace 6.25\%/0.20\%/4.55 & \enspace 9.72\%/0.08\%/2.79 & \textbf{16.67\%}/\textbf{0.05\%}/\textbf{2.00} \\
             & CodeT5 & 15.07\%/\textbf{0.02\%}/1.83 & 14.38\%/0.11\%/3.98 & \enspace 8.90\%/0.18\%/3.06 & 11.64\%/0.09\%/1.63 & \textbf{15.75\%}/0.05\%/\textbf{1.54} \\
            \hline
            \multirow{3}{*}{\makecell[c]{Functionality \\ Classification \\ C++1000}} 
             & CodeBERT & 19.44\%/0.24\%/1.59 & \enspace 9.16\%/0.31\%/3.29 & \enspace 6.45\%/0.66\%/3.10 & 12.68\%/0.17\%/1.32 & \textbf{20.19\%}/\textbf{0.08\%}/\textbf{1.31} \\
             & GraphCodeBERT & \enspace 1.56\%/0.33\%/2.78 & \enspace 3.49\%/0.34\%/4.51 & \enspace 2.61\%/0.43\%/4.57 & \enspace 3.67\%/0.31\%/2.52 & \enspace \textbf{5.12\%}/\textbf{0.06\%}/\textbf{2.32} \\
             & CodeT5 & 14.20\%/0.25\%/1.62 & \enspace 7.01\%/0.23\%/2.22 & \enspace 5.17\%/0.30\%/2.79 & \enspace 9.57\%/0.11\%/1.35 & \textbf{14.41\%/}\textbf{0.05\%}/\textbf{1.17} \\
            \hline
            \multirow{3}{*}{\makecell[c]{Functionality \\ Classification \\ Python800}} 
             & CodeBERT & 26.01\%/0.16\%/2.46 & 14.76\%/0.20\%/3.73 & 11.75\%/0.26\%/2.20 & 16.71\%/0.14\%/2.21 & \textbf{27.34\%}/\textbf{0.07\%}/\textbf{1.56} \\
             & GraphCodeBERT & 16.31\%/0.08\%/3.14 & 12.40\%/0.09\%/4.43 & \enspace 9.30\%/0.14\%/3.63 & 12.40\%/0.09\%/4.21 & \textbf{17.79\%}/\textbf{0.06\%}/\textbf{1.41} \\
             & CodeT5 & 17.69\%/0.16\%/2.43 & 14.27\%/0.11\%/3.08 & 11.47\%/0.20\%/2.08 & 14.45\%/0.07\%/2.22 & \textbf{18.48\%}/\textbf{0.03\%}/\textbf{1.80} \\
            \hline
            \multirow{3}{*}{\makecell[c]{Functionality \\ Classification \\ Java250}} 
             & CodeBERT & 23.42\%/0.07\%/1.89 & 18.92\%/0.22\%/3.47 & 17.30\%/0.61\%/3.66 & 21.80\%/0.24\%/1.68 & \textbf{29.01\%}/\textbf{0.06\%}/\textbf{1.23} \\
             & GraphCodeBERT & 18.73\%/0.31\%/1.90 & 16.31\%/0.35\%/3.97 & \enspace 9.37\%/0.65\%/4.02 & 19.34\%/0.40\%/1.74 & \textbf{25.68\%}/\textbf{0.27\%}/\textbf{1.73} \\
             & CodeT5 & 22.75\%/0.03\%/2.09 & 11.38\%/0.11\%/4.09 & \enspace 7.67\%/0.27\%/3.30 & 16.67\%/0.15\%/2.13 & \textbf{25.93\%}/\textbf{0.03\%}/\textbf{1.57} \\
            \midrule
            \multicolumn{2}{c}{Average} & 16.91\%/0.52\%/2.25 & 14.65\%/0.41\%/3.79 & 10.84\%/0.52\%/3.27 & 15.50\%/0.34\%/2.27 & \textbf{21.91\%}/\textbf{0.09\%}/\textbf{1.58} \\
            \bottomrule
        \end{tabular}
    \end{adjustbox}
\end{table*}

\begin{tcolorbox}[]
\textbf{Answer to RQ2:} All the three components, i.e., randomized-smoothing-based mispredicted input identification, attention-based noise localization, and MCIP-based noise cleansing, make contributions to the overall effectiveness and efficiency of \tech{}, demonstrating the necessity of each of them.
\end{tcolorbox}

\subsection{RQ3: Influence of Hyper-parameters}
\label{sec:rq3}

\begin{table}[]
    \caption{Influence of hyper-parameter $\theta$ in terms of CSR ($\uparrow$)/MCR ($\downarrow$)/the average time spend on denoising an input ($\downarrow$)}
    \label{tab:theta}
    \centering
    \tabcolsep=2.5mm
    \begin{adjustbox}{max width=0.50\textwidth,center}
        \begin{tabular}{ cccccc }
            \toprule
        	\textbf{$\theta$} & \textbf{1} & \textbf{2} & \textbf{3} & \textbf{4} & \textbf{5} \\
        	\midrule
                CSR ($\uparrow$) & 21.91\% & 22.85\% & 23.95\% & 25.27\% & 26.08\%  \\ 
                MCR ($\downarrow$) & 0.09\% & 0.14\% & 0.16\% & 0.20\% & 0.29\%  \\ 
                Time (s) ($\downarrow$) & 0.48 & 0.63 & 1.03 & 1.43 & 1.70  \\ 
            \bottomrule
        \end{tabular}
    \end{adjustbox}
\end{table}

\begin{table}[]
    \caption{Influence of hyper-parameter $N$ in terms of CSR ($\uparrow$)/MCR ($\downarrow$)/the average time spend on denoising an input ($\downarrow$)}
    \label{tab:n}
    \centering
    \tabcolsep=2.5mm
    \begin{adjustbox}{max width=0.50\textwidth,center}
        \begin{tabular}{ cccccc }
            \toprule
        	\textbf{$N$} & \textbf{$\times$1} & \textbf{$\times$2} & \textbf{$\times$3} & \textbf{$\times$4} & \textbf{$\times$5} \\
        	\midrule
        	CSR ($\uparrow$) & 21.91\% & 23.30\% & 24.66\% & 25.25\% & 25.99\%  \\ 
                MCR ($\downarrow$) & 0.09\% & 0.09\% & 0.08\% & 0.08\% & 0.08\%  \\ 
                Time (s) ($\downarrow$) & 0.48 & 0.71 & 0.87 & 1.13 & 1.63  \\ 
            \bottomrule
        \end{tabular}
    \end{adjustbox}
\end{table}

\subsubsection{Setup}
We studied the influence of two hyper-parameters in \tech{} (i.e., $\theta$ and $N$ introduced in Section~\ref{sec:mispredicted_input_identification}). 
By default, we set $\theta$ to 1 and $N$ to the number of the identifiers in the given input.
We investigated the effectiveness and efficiency of \tech{} under different settings, i.e., $\theta=\{1,2,3,4,5\}$ and $N$ is $\{1,2,3,4,5\}$ times the number of identifiers in a given code snippet.
When changing the setting of one hyper-parameter, the setting of another hyper-parameter is set to the default value.

\subsubsection{Results}
Table~\ref{tab:theta} and Table~\ref{tab:n} show the results of \tech{} under different hyper-parameter settings in terms of average CSR, average MCR, and the average time spent on denoising an input code across all the subjects.
We found that with the setting of $\theta$ increasing, more mispredicted inputs can be identified, leading to larger CSR; meanwhile more correctly-predicted inputs are identified as mispredicted ones due to making larger perturbations, leading to larger MCR.
We need to balance CSR and MCR for $\theta$.
With the setting of $N$ increasing, the randomized-smoothing-based strategy can make more accurate identification for mispredicted inputs, leading to larger CSR and smaller MCR.
However, larger settings of $N$ (as well as $\theta$) can incur more time cost due to generating more perturbed inputs and involving more invocations to the target code model.
For on-the-fly improving (deployed) model performance, high efficiency is quite critical.
Hence, by balancing the effectiveness (in terms of different metrics) and efficiency of \tech{}, we set $\theta$ to 1 and $N$ to $\times$1 (the number of identifiers in a given code snippet) as the default configurations in \tech{} for practical use.


\begin{tcolorbox}[]
\textbf{Answer to RQ3:} As the settings of $\theta$ and $N$ increase, \tech{} is more effective in terms of CSR, but less efficient in terms of the average time spent on denoising an input (and less effective in terms of MCR for larger $\theta$). 
We obtained their default settings by balancing effectiveness and efficiency for practical use.
\end{tcolorbox}
\section{Discussion}
\label{sec:discussion}

\subsection{Future Work}
\tech{} is the first technique to on-the-fly improve the performance of deep code models (in deployment) through input denoising.
Its effectiveness and efficiency have been demonstrated in our empirical study, but there are some aspects in \tech{} that can be further extended.

(1) The current \tech{} denoises inputs at the identifier level.
In fact, noise may exist in other types of code features, such as code structure~\cite{henke2022semantic,zhang2022towards}, which cannot be handled by the current \tech{}.
For example, structure-level noise can largely affect the code (involving more code elements), making it challenging to achieve effective denoising via minor identifier-level denoising.
In the future, we can further enhance \tech{} by localizing and cleansing more types of noise.
(2) According to our results, \tech{} still makes some correct predictions become mispredictions due to imperfect identification of mispredicted inputs.
In the future, we can design more accurate methods (e.g. constructing a classification model) to identify mispredicted inputs in order to improve the overall effectiveness of \tech{}.
In particular, we conducted a preliminary study to explore the improvement brought by perfect mispredicted input identification on the three code models with the Defect Prediction dataset.
That is, we directly used the ground-truth mispredicted inputs as the input of the subsequent noise localization and noise cleansing components.
The results show that it successfully corrects 40.67\% mispredictions on average, demonstrating a large improvement compared to \tech{} (21.12\%).
Note that \tech{} can achieve better performance if achieving perfect noise localization and cleansing as well.
(3) We can extend the usage scenario of \tech{}.
Currently, \tech{} has been demonstrated effective to denoise natural inputs, but in fact it can be used to defend against adversarial attacks.
In particular, we conducted a preliminary study to investigate the effectiveness of \tech{} to defend against adversarial examples (generated by the state-of-the-art adversarial attack technique ALERT~\cite{yang2022natural}) on the three code models with the Defect Prediction dataset.
The results show that \tech{} identifies 90.59\% of adversarial examples, and successfully denoises 56.17\% of them.
(4) \tech{} remains ineffective for some code snippets.
In the future, we intend to conduct a manual analysis on representative examples. 
It may help identify common characteristics of noisy identifiers, thereby contributing to further improvement.
(5) Depending on internal attention weights for localizing noisy identifiers within \tech{} could limit its generality, especially for the cases where deep code models do not use attention mechanisms or they are in black-box scenarios.
In the future, we will explore a more general method of localizing noisy identifiers, thus broadening its usage scope.

\subsection{Orthogonality between \tech{} and Fine-tuning}
In fact, \tech{} and the fine-tuning strategy are orthogonal to a large extent.
On the one hand, after fine-tuning the deep code model, \tech{} can be still applied to further improve the performance of the fine-tuned model.
On the other hand, \tech{} can be a data augmentation method for the fine-tuning strategy, since the inputs generated by \tech{} are much cleaner code snippets.
In the future, we will study the combination effect of the two techniques.

\subsection{Discussion about Randomized Smoothing}
Intuitively, randomized smoothing can be directly used to improve model performance.
Given an input image, randomized smoothing randomly constructs a set of perturbed images and adopts the majority predicted class on them as the final prediction result, instead of the original prediction result of the input image.
This practice has been discussed and investigated in the field of image processing~\cite{kumar2020certifying,mehra2021robust,chen2022input}.
However, its effectiveness is unsatisfactory due to ignoring much information (such as prediction confidence information)~\cite{kumar2020certifying}.

In \tech{}, we just used randomized smoothing to identify mispredicted inputs if the original prediction result is different from the majority predicted class, rather than directly use the majority predicted class as the final one.
The former is less aggressive than the latter, and then the follow-up components of noise localization and noise cleansing in \tech{} can utilize more information to obtain the correct prediction result more effectively.

In particular, we conducted a preliminary study to investigate the effectiveness of directly using the majority predicted class as the final one for each identified mispredicted input.
We used the three deep code models with the Defect Prediction dataset as the representative.
The results show that \tech{} has much larger CSR values than this method (20.92\% vs 1.47\%), and has smaller MCR values than the latter (0.28\% vs 0.33\%), demonstrating the effectiveness of applying randomized smoothing to mispredicted input identification (instead of directly applying it to inferring final prediction).

\subsection{Threats to Validity}
\label{sec:threats}

The threat to \textit{internal} validity mainly lies in the implementation of \tech{} and experimental scripts. 
To address this threat, we used some mature tool (i.e., tree-sitter), 
carefully conducted testing and code review, and released the artifact for replication and practical use.

The threat to \textit{external} validity mainly lies in the used subjects. 
We reduced this threat by adopting three most popular pre-trained code models and six widely-used datasets.
In particular, we carefully considered the diversity of our subjects as presented in Section~\ref{sec:sub}.
Our results demonstrate the stable superiority of \tech{} over baselines or variants across all the 18 deep code models.
According to our results (that are put at the homepage), the effectiveness of \tech{} is also not affected by different lengths of code snippets.


\section{Related Work}
\label{sec:related}


To improve the performance of deep code models, there are two main methods, i.e., designing more advanced neural networks for retraining models and incorporating more data for fine-tuning models.

Regarding the former, 
Bielik et al.~\cite{bielik2020adversarial} applied (m+1)-class to the deep code model and then retrained the model to allow it to abstain the predictions on unconfident inputs.
Bui et al.~\cite{bui2021treecaps} incorporated tree-based convolutional neural networks into capsule networks for better learning of code on abstract syntax trees, in order to improve the model performance on program comprehension tasks.
Recenly, some work adopted contrastive learning to improve the deep code model performance based on the new auxiliary networks and loss functions~\cite{wang2021syncobert,jia2022clawsat,liu2023contrabert,wang2022heloc}.

Regarding the latter, some existing work proposed to construct adversarial examples (i.e., code snippets) to fine-tune a deep code model in order to improve model performance~\cite{zhang2022towards,yefet2020adversarial}.
Similarly, Yu et al.~\cite{yu2022data}, and Allamanis et al.~\cite{allamanis2021self} designed program transformation rules to augment code data for improving code model generalization. 
Mi et al.~\cite{mi2021effectiveness} generated additional data from auxiliary generative adversarial networks by treating code snippets as images.
Recently, Wang et al.~\cite{wang2022bridging} incorporated curriculum learning to augment code data in order to optimize the code model fine-tuning process,
which can reduce overfitting to the target dataset and thus improve model robustness.

However, both kinds of techniques are not applicable to on-the-fly improve performance of deep code models in deployment, which is the main target of \tech{}.
Different from them, \tech{} neither retrains nor fine-tunes the target deep code model, but improves the model performance at the input code level through input denoising.
Our study also demonstrated better effectiveness and efficiency of \tech{} compared with the widely-used fine-tuning strategy for improving model performance.
Moreover, these existing techniques are mostly inexplicable for improving model performance, but \tech{} can compare original inputs and corresponding denoised inputs to facilitate the understanding of correcting mispredictions.



Besides, there are some techniques that improve model performance at the input level in the field of Computer Vision.
For example, 
Xiao et al.~\cite{xiao2021self} proposed SelfChecker to monitor model outputs using internal layer features.
It provides an alarm and an alternative prediction if the internal layer features of the model are inconsistent with the final prediction. 
However, SelfChecker 
is specific to Convolutional Neural Network and Computer Vision tasks due to relying on Grad-CAM~\cite{selvaraju2017grad}. 
Xiao et al.~\cite{xiao2022repairing} proposed InputReflector, which uses two auxiliary models and similar training data to replace mispredicted images caused by deviation and out-of-distribution.
Here, InputReflector applied six image transformation rules to construct in-distribution data.
Due to the discrete space, complex constraints in source code, and the inherent nature of deep code models, all of these techniques specific to images cannot be applied to deep code models. 

There are also some work on ensuring the quality of deep learning systems from the testing perspective, which involve different levels such as model-level~\cite{you2023regression,gerasimou2020importance,wang2020dissector,stocco2020misbehaviour}, program-level~\cite{yan2021exposing,zhang2018empirical}, library-level~\cite{wang2020deep,chen2022toward}, and compiler-level~\cite{ma2023fuzzing,shen2021comprehensive} testing.

\section{Conclusion}
To improve performance of deep code models, the existing techniques are mainly based on retraining or fine-tuning.
However, these techniques cannot on-the-fly improve performance of (deployed) models, and are time-consuming and inexplicable.
Input denoising is effective to overcome these limitations, but the characteristics of deep code models (discrete code space and complex syntactic and semantic constraints) make the input denoising techniques in other fields not applicable to deep code models.
In this work, we propose the first input denoising technique (\tech{}) for on-the-fly improve performance of deep code models.
It consists of randomized-smoothing-based mispredicted input identification, attention-based noise localization, and MCIP-based noise cleansing.
Our extensive study on 18 deep code models demonstrates the effectiveness and efficiency of \tech{}, significantly outperforming the widely-used fine-tuning technique.

\section*{Acknowledgements}
This work was supported by the National Natural Science Foundation of China Grant Nos. 62322208, 62002256, and CCF Young Elite Scientists Sponsorship Program (by CAST),
and NSF Nos. 1901242, 1910300.

\balance
\bibliographystyle{IEEEtrans}
\bibliography{reference}

\end{document}